\documentstyle[prl,aps,preprint,tighten,floats]{revtex}
\begin{document}
\draft
\preprint{UPR-674-T, NSF-ITP-95-65, hep-th \# 9507160}
\date{July 1995}
\title{Singular BPS Saturated States and Enhanced Symmetries of
Four-Dimensional N=4 Supersymmetric String Vacua }
\author{Mirjam Cveti\v c
\thanks{E-mail address: cvetic@cvetic.hep.upenn.edu}$^{1,2}$
and Donam Youm\thanks{E-mail address: youm@cvetic.hep.upenn.edu}$^1$}
\address {$^1$ Department of Physics and Astronomy \\
          University of Pennsylvania, Philadelphia PA 19104-6396\\
          and\\  $^2$ Institute for Theoretical Physics,\\
          University of California, Santa Barbara, CA 93106-4030}
\maketitle
\begin{abstract}
{A class of supersymmetric (BPS saturated), static, spherically
symmetric solutions of four-dimensional effective $N=4$ supersymmetric
superstring vacua, which become massless at special points of moduli
space, is studied in terms of the fields of the effective heterotic
string theory compactified on a six-torus.  Those are singular
four-dimensional solutions corresponding to $O(6,22,Z)$ orbits of
dyonic configurations (with zero axion), whose left-moving as well as
right-moving electric and magnetic charges are orthogonal (light-like
in the $O(6,22,Z)$ sense), while the $O(6,22,Z)$ norms of both the
electric and magnetic charges are negative.  Purely electric [or purely
magnetic] and dyonic configurations preserve $1\over 2$ and $1\over 4$
of $N=4$ supersymmetry, respectively, thus belonging to the vector and
the highest spin $3\over 2$ supermultiplets, respectively.   Purely
electric [or purely magnetic] solutions (along with an infinite tower of
$SL(2,Z)$ transformed states) become massless at a point of the
corresponding ``one-torus'', thus  may contribute to the enhancement
of non-Abelian gauge symmetry, while dyonic solutions become
simultaneously massless at a point of the corresponding two-torus,
and thus may in addition contribute to the enhancement of the local
supersymmetry there.}
\end{abstract}

Recently, it was recognized \cite{HTI,WITTENII} that non-trivial
configurations, {\it e.g.}, Bogomol'yni-Prasad-Sommerfield (BPS)
saturated states, play a crucial role in addressing the full
non-perturbative dynamics of string theory.  When such configurations
become light they can affect the low energy dynamics of the string
theory in a crucial way \cite{Hull,STROM,HTII}.

Within Type IIA string compactified on Calabi-Yau manifolds, it was
pointed out by Strominger \cite{STROM} and further studied in Refs.
\cite{GRMOSTR,WITTENI} that massless supersymmetric black holes
play a crucial role in the full string theory dynamics at the
conifold points of moduli space.

Within four-dimensional $N=4$ superstring vacua, Hull and Townsend
\cite{Hull,HTII} show that at special points of moduli space massless
BPS saturated states can occur, contributing to a phenomenon which
is a generalization of the Halpern-Frenkel-Ka\v c (HFK) mechanism;
namely, at special points of moduli space along with the perturbative
electrically charged massless string states, which enhance the gauge
symmetry to the non-Abelian one, there are massless BPS saturated
magnetic monopoles and a tower of $SL(2,Z)$ related BPS saturated
dyons which contribute to the new phase of the enhanced non-Abelian
gauge symmetry.

In somewhat parallel developments, Behrndt \cite{BEHRNDT} found
four-dimensional, electrically charged supersymmetric black holes,
which can become massless at special points of moduli space.
They correspond to exact string backgrounds which are obtained by
 dimensionally reducing  supersymmetric
plane-wave solutions of effective ten-dimensional heterotic
string theory.  These solutions were generalized to the corresponding
multi-black holes by  Kallosh \cite{KALL1}, while the corresponding
exact magnetic solutions and the physical implications of
the whole class of these solutions were studied by Kallosh and
Linde \cite{KALLIND}.

The purpose of the paper is to study properties of a general class
of BPS saturated, spherically symmetric solutions of effective
four-dimensional $N=4$ supersymmetric string vacua, which become
massless at special points of moduli space.  We  parameterize
these solutions in terms of fields of the effective heterotic string
theory compactified on a six-torus.  BPS saturated solutions in the
effective Type II string compactified on $K_3\times T^2$ surface are
related to those of the toroidally compactified heterotic string
through field redefinitions, since both of the effective actions
in four dimensions are the same.  A set of BPS states in one
theory may turn out to be identified with elementary string states of
a dual theory, thus providing further evidence for the conjectured
duality between the two theories, whose origin lies in the
string-string duality conjecture \cite{DUFFSS,WITTENII,HTI,STRDUAL}
of the corresponding six-dimensional theories.

The effective field theory of massless bosonic fields for the
heterotic string on a Narain torus \cite{NARAIN} at a generic point
of moduli space is obtained by compactifying the ten-dimensional
$N=1$ Maxwell/Einstein supergravity theory on a six-torus
\cite{SCHWARZ,SEN2}.  The effective four-dimensional action
\footnote{See Refs. \cite{SCHWARZ,SEN2} for notational conventions and
the relationship of four-dimensional fields to the corresponding
ten-dimensional ones.  Also, since we are studying the semiclassical
configurations of the effective action we are not addressing
$\alpha'$ corrections.}
for massless bosonic fields consists of the graviton $g_{\mu\nu}$, 28
$U(1)$ gauge fields ${\cal A}^i_{\mu} \equiv (A^{(1)\, m}_{\mu},
A^{(2)}_{\mu\, m}, A^{(3)\, I}_{\mu})$, corresponding to the gauge fields
of dimensionally reduced  ten-dimensional metric (Kaluza-Klein sector),
two-form fields, and Yang-Mills fields, respectively, and 134 scalar
fields.  The scalar fields consist of the dilaton $\phi$,
which parameterizes the strength of the string coupling, the axion field
$\Psi$, which is obtained from the two-form field $B_{\mu\nu}$ through
the duality transformation, and a symmetric $O(6,22)$ matrix $M$ of
132 scalar fields.  The matrix $M$ consists of 21 internal metric $g_{mn}$
components, 15 pseudo-scalar fields $B_{mn}$, and 96 scalar fields
$a^I_m$, which arise from the dimensionally reduced ten-dimensional
metric, two-form field and Yang Mills fields, respectively.  Here,
$(\mu,\nu)=0,\cdots,3$, $(m,n)=1,\cdots, 6$ and $I=1,\cdots,16$.

The classical four-dimensional effective action is invariant under
the $O(6,22)$ transformations ($T$-duality) \cite{SCHWARZ,SEN2}:
\begin{equation}
M \to \Omega M \Omega^T ,\ \ \ {\cal A}^i_{\mu} \to \Omega_{ij}
{\cal A}^j_{\mu}, \ \ \ g_{\mu\nu} \to g_{\mu\nu}, \ \ \ \phi \to \phi .
\label{tdual}
\end{equation}
Here, $\Omega \in O(6,22)$, {\it i.e.}, $\Omega^T L \Omega = L$,
where $L$ is an $O(6,22)$ invariant matrix.  In addition, the
corresponding equations of motion and Bianchi identities have the
invariance under the $SL(2,R)$ transformations ($S$-duality)
\cite{SEN2,STRWK}:
\begin{equation}
S \to S^{\prime}={{aS+b}\over{cS+d}},\ M\to M ,\ g_{\mu\nu}\to
g_{\mu\nu},\ {\cal F}^i_{\mu\nu} \to {\cal F}^{\prime\, i}_{\mu\nu} =
(c\Psi + d){\cal F}^i_{\mu\nu} + ce^{-\phi} (ML)_{ij}
\tilde{\cal F}^j_{\mu\nu},
\label{sdual}
\end{equation}
where $S \equiv \Psi + i e^{-\phi}$, $\tilde{\cal F}^{i\,\mu\nu} =
{1\over 2}(\sqrt{-g})^{-1} \varepsilon^{\mu\nu\rho\sigma}
{\cal F}^i_{\rho\sigma}$, and $a,b,c,d \in I\!\!R$ satisfy $ad-bc=1$.
By embedding an Abelian gauge group in a non-Abelian one,
one can see that the instanton effect breaks the $SL(2,R)$ symmetry
down to $SL(2,Z)$, referred to as $S$-duality.  The allowed discrete
electric $\vec{Q}$ and magnetic $\vec{P}$ charges are determined
\cite{SEN2} by $T$- and $S$- duality constraints of the toroidally
compactified heterotic string \cite{NARAIN} and
Dirac-Schwinger-Zwanzinger-Witten (DSZW) quantization condition
\cite{DSZ,WITTENIII}; both of the ``lattice charge vectors''
\cite{SEN2}, $\vec{\beta} \equiv L\vec{P}$ and $\vec{\alpha} \equiv
e^{-\phi_{\infty}}M^{-1}_{\infty}\vec{Q}-\Psi_{\infty}
\vec{\beta}$, lie in an even, self-dual, Lorenzian lattice $\Lambda$
with signature $(6,22)$. The world-sheet instanton effects break
$O(6,22,R)$ invariance of the effective action down to its discrete
subgroup $O(6,22,Z)$ referred to as $T$-duality. $T$-duality is an exact
string symmetry to all orders in string perturbation and is assumed to
survive non-perturbative corrections.

For an arbitrary asymptotic value $M_{\infty}$ of the moduli
fields, one can perform \cite{SEN2} the following simultaneous
$O(6,22,R)$ rotations on the lattice $\Lambda$ and the matrix
$M_{\infty}$: $M_{\infty} \to \hat{M}_{\infty} =
\Omega M_{\infty} \Omega^T$ and $\Lambda \to \hat{\Lambda} = L
\Omega L \Lambda$ ($\Omega \in O(6,22,R)$), in such a way that
$\hat{M}_{\infty}= I_{28}$.  The charge configuration is now
described by a new lattice $\hat{\Lambda}$.  The subset of $O(6,22,Z)$
transformations that preserves this new asymptotic value of
${\hat M}_{\infty}$ is $O(6,Z)\times O(22,Z)$.  Additionally, one
imposes $SL(2,R)$ transformation to bring $S_{\infty} \to
\breve{S}_{\infty} = i$.  The subset of $SL(2,Z)$ transformations
that preserves the asymptotic value $\breve{S}_{\infty} = i$ is
$SO(2,Z)$.  And the charge configuration is described by a new lattice
$\breve{\Lambda}$.  After imposing a subset of $O(6,Z)\times O(22,Z)$
and $SO(2,Z)$ transformations to generate a new solution from the
generating solution, one has to undo the above $O(6,22,R)$ and $SL(2,R)$
rotations in order to obtain a new configuration with an arbitrary
asymptotic value of the moduli field specified by $M_{\infty}$
and an arbitrary asymptotic value $S_{\infty}$ of the complex scalar.

The spectrum of BPS saturated (supersymmetric), static, spherically
symmetric configurations at generic points of moduli space of
toroidally compactified heterotic string is both $O(6,22,Z)$ and
$SL(2,Z)$ invariant.  In Ref. \cite{CVETICYOUM} we derived the explicit
form of a general class of such configurations with 56 charges subject
to one constraint.  It corresponds to the states which can be obtained
by $SL(2,Z)$ transformations on configurations {\it  with zero axion and
the most general allowed dyonic charges}.  The latter set of
configurations corresponds to $O(6,22,Z)$ orbits of dyonic configurations
(with zero axion), whose left-moving as well as right-moving electric
and magnetic charges are orthogonal, {\it i.e.}, light-like in the
$O(6,22,Z)$ sense.

The BPS states with general charge configurations can be obtained from
the generating solutions through the $O(6,22,Z)$ and the $SL(2,Z)$
transformations.   It turns out \cite{CVETICYOUM} that the generating
solution corresponds to the one where from scalar fields only the
diagonal components $g_{mm}$ of the internal metric and the dilaton
field $\phi$ are turned on.  The charge lattice can be decomposed
into sublattices which are $O(6,22,Z)$ orbits, each of which having
a fixed value of the $O(6,22,Z)$ norm for the charge vectors.
All the BPS states related through the $O(6,22,Z)$ and $SL(2,Z)$
transformations have the same thermal and space-time properties.

For the purpose of illustrating the symmetry enhancement
at particular points of moduli space, we shall concentrate on the
generating solution, and choose $M_{\infty}$ and $S_\infty$
to be diagonal and purely imaginary, respectively.
Other solutions with more general charge configurations and different
choices of $M_{\infty}$ and $S_\infty$ allow for a larger enhancement
of gauge symmetry.

Now, we recapitulate the results of Ref. \cite{CVETICYOUM} for the
explicit form of the generating solution along with the above choice of
$M_{\infty}$ and $S_\infty$.  It is parameterized by two magnetic and
two electric charges with electric and magnetic charges arising from
different $U(1)$ groups; the two magnetic [electric]
charges arise from the Kaluza-Klein sector gauge field $A^{(1)\, m}_{\phi}$
[$A^{(1)\, n}_{t}$] and the corresponding two-form $U(1)$ field
$A^{(2)}_{\phi\, m}$ [$A^{(2)}_{t\, n}$].  Here, $m\ne n$.  Without loss
of generality, we choose the non-zero charges to be $P^{(1)}_1,
P^{(2)}_1, Q^{(1)}_2, Q^{(2)}_2$.

The upper $\varepsilon_{u}$ and lower $\varepsilon_{\ell}$ two-components
of the four-component Killing spinors are subject to the constraints:
$\Gamma^1\, \varepsilon_{u,\ell}=i\eta_P\,\varepsilon_{\ell,u}$
if ${ P}^{(1)}_1 \ne 0 $ and/or ${P}^{(2)}_1 \ne 0$, and
$\Gamma^2\,\varepsilon_{u,\ell}=\mp \eta_Q\, \varepsilon_{\ell,u}$ if
${ Q}^{(1)}_2 \ne 0 $ and/or ${ Q}^{(2)}_2 \ne 0$.  Here,
$\eta_P$ and $\eta_Q$ are  $\pm 1$ and $\Gamma^{1,2}$ are the gamma
matrices of the corresponding $SO(6)$ Clifford algebra.  Non-zero
magnetic and electric charges each break $1\over 2$ of the remaining
supersymmetries.  Thus, purely electric [or purely magnetic]
configurations preserve $1\over 2$, while dyonic solutions
preserve $1\over 4$, of $N=4$ supersymmetry in four dimensions.
The first and the second sets of configurations fall into vector-
and hyper-supermultiplets with highest spins 1 and $3\over 2$
\footnote{Massive highest spin $3\over 2$ multiplets were addressed
in the context of supergravity in Refs. \cite{STRATH,FERRARA} and in
the context of massive BPS supermultiplets by Kallosh \cite{KALL3}.},
respectively.

With the static, spherically symmetric Ansatz for the four-dimensional
space-time metric in the Einstein frame: $g_{\mu\nu} dx^{\mu} dx^{\nu} =
\lambda(r) dt^2 - \lambda^{-1}(r)dr^2 - R(r)(d\theta^2 + {\rm sin}^2
\theta d\phi^2)$, and with the internal diagonal metric $g_{mn} =
\delta_{mn}g_{mm}$ as well as the dilaton $\phi$ depending only on the
radial coordinate $r$, the explicit form for the solution
is given by \cite{CVETICYOUM}:
\begin{eqnarray}
\lambda &=& r^2/[(r-\eta_P{\bf P}^{(1)}_{1\,\infty})
(r-\eta_P {\bf P}^{(2)}_{1\,\infty})
(r- \eta_Q {\bf Q}^{(1)}_{2\,\infty})
(r-\eta_Q {\bf Q}^{(2)}_{2\,\infty})]^{1\over 2},
\nonumber\\
R &=& [(r-\eta_P {\bf P}^{(1)}_{1\,\infty})
(r - \eta_P {\bf P}^{(2)}_{1\,\infty})
(r - \eta_Q {\bf Q}^{(1)}_{2\,\infty})
(r-\eta_Q {\bf Q}^{(2)}_{2\,\infty})]^{1\over 2},
\nonumber\\
e^{\phi}&=&e^{\phi_{\infty}}\left [{(r-\eta_P {\bf P}^{(1)}_{1\,\infty})
(r- \eta_P {\bf P}^{(2)}_{1\,\infty})} \over
{(r- \eta_Q {\bf Q}^{(1)}_{2\,\infty})
(r- \eta_Q {\bf Q}^{(2)}_{2\,\infty})}\right]^{1\over 2},
\nonumber\\
g_{11}&=&g_{11\,\infty}\left ({{r- \eta_P {\bf P}^{(2)}_{1\,\infty}} \over
{r-\eta_P {\bf P}^{(1)}_{1\,\infty}}} \right ), \
g_{22}=g_{22\,\infty}\left ({{r- \eta_Q {\bf Q}^{(1)}_{2\,\infty}} \over
{r- \eta_Q {\bf Q}^{(2)}_{2\,\infty}}}\right ),\
g_{mm}=g_{mm\,\infty}\ \ \ (m \neq 1,2).
\label{gensol}
\end{eqnarray}
Here, the radial coordinate is chosen so that the horizon is at $r=0$.
The properties of the generating solution depend only on the four
screened charges $({\bf  P}^{(1)}_{1\,\infty}, {\bf P}^{(2)}_{1\,\infty},
{\bf Q}^{(1)}_{2\,\infty}, {\bf Q}^{(2)}_{2\,\infty}) \equiv
e^{-{\phi_{\infty}\over 2}}(g^{1\over 2}_{11\,\infty}P_1^{(1)},
g^{-{1\over 2}}_{11\,\infty}P_1^{(2)}, g^{-{1\over 2}}_{22\,\infty}
Q_2^{(1)}, g^{1\over 2}_{22\,\infty}Q_2^{(2)}) = (e^{-{\phi_{\infty}
\over 2}}g^{1\over 2}_{11\,\infty}\beta_1^{(2)}, e^{-{\phi_{\infty}
\over 2}} g^{-{1\over 2}}_{11\,\infty}\beta_1^{(1)}, e^{{\phi_{\infty}
\over 2}} g^{{1\over 2}}_{22\,\infty}\alpha_2^{(1)}, e^{{\phi_{\infty}
\over 2}}g^{-{1\over 2}}_{22\,\infty}\alpha_2^{(2)})$, where
the  lattice charge vectors  $\vec\alpha$ and $\vec\beta$ lie on
the lattice $\Lambda$.

The requirement that the ADM mass of the above configuration saturates
the Bogomol'ny bound requires the choice of parameters $\eta_{P,Q}$
to be such that $\eta_P\, {\rm sign}(P_1^{(1)}+P_1^{(2)})= -1$ and
$\eta_Q\, {\rm sign}(Q_2^{(1)}+Q_2^{(2)})=-1$, thus yielding the
positive semi-definite ADM mass of the following form
\footnote{The $SL(2,Z)$ and $O(6,22,Z)$ invariant form of ADM mass,
which generalizes the Bogomol'nyi mass formula to the case of BPS
states preserving $1\over 4$ of the original supersymmetry, only, is
given in Ref. \cite{CVETICYOUM}.}:
\begin{equation}
M_{\rm BPS} = e^{-{\phi_{\infty} \over 2}}|g^{1\over 2}_{11\,\infty}
\beta^{({2})}_1 + g^{-{1\over 2}}_{11\,\infty}\beta^{(1)}_1| +
e^{{\phi_\infty}\over 2}|g^{{1\over 2}}_{22\,\infty}\alpha^{(1)}_2+
g^{-{1\over 2}}_{22\,\infty}\alpha^{(2)}_2|.
\label{ADMmass}
\end{equation}

Note that at generic points of moduli space, in Ref. \cite{CVETICYOUM}
we studied regular BPS saturated solutions.
In this case, the requirement of the absence of naked singularities
restricts the relative signs of the two magnetic [and two electric]
charges to be the same.  For such solutions, $r=0$ corresponds to
the Reissner-Nordstr\" om-type horizon,  null singularity, and
naked singularity when four charges, three [or two] charges,
and one charge are nonzero, respectively.

On the other hand, Eq. (\ref{gensol}) implies {\it when the relative
signs for the two magnetic [and/or two electric] charges are opposite
the solutions are always singular}.  When both the magnetic and the
electric charges have opposite relative signs, the curvature singularity
takes place at $r=r_{\rm sing}\equiv{\rm max}\{{\rm min}
[|{\bf P}^{(1)}_{1\,\infty}|,|{\bf P}^{(2)}_{1\,\infty}|],
{\rm min}[|{\bf Q}^{(1)}_{2\,\infty}|,|{\bf Q}^{(2)}_{2\,\infty}|]\} >0$.
In addition, these configurations, in turn, become massless at special
points of moduli space; when the magnitudes of the two screened electric
[and two magnetic] charges of the generating solution are equal,
the corresponding BPS saturated solutions in this class have zero ADM
mass for special values of $M_{\infty}$.

A class of such purely electrically charged configurations is related
to the solution, recently found by Behrndt \cite{BEHRNDT,BEHRNDT2},
which was obtained by dimensionally reducing  supersymmetric gravitational
waves of the effective ten-dimensional heterotic string theory.
Generalizations to the corresponding multi-black hole solutions
and the corresponding exact (in $\alpha'$ expansion) magnetic
solutions were given by Kallosh \cite{KALL1}, and by Kallosh and Linde
\cite{KALLIND}, respectively.  In the latter work, the physical
properties of such configurations were further addressed;
they repel massive particles.

Just as in the case of purely electric [or purely magnetic] singular
solutions \cite{KALLIND}, a general class of such dyonic configurations,
when the magnetic and/or the electric charges have opposite relative signs,
again repel massive particles.  In particular, the traversal time of
the geodesic motion for a test particle with energy $E$, mass $m$ and
zero angular momentum along the radial coordinate $r$, as measured by
an asymptotic observer, can be written as
\footnote{Note that for the general class of singular solutions
studied here, $\lambda \ge 1$ for $r$ small enough, while for the
regular solutions, studied in Ref. \cite{CVETICYOUM} (with the same
relative signs for the corresponding charges), $\lambda\le 1$.  Thus,
for the latter type of configurations the particles are always
attracted toward the singularity.  In the case of only one non-zero
charge, the regular solution has a naked singularity at $r=0$,
{\it i.e.}, $t(r=0)$ is finite.}:
\begin{equation}
t(r)=\int^r_{r_{\infty}} {{E\,dr}\over{\lambda(r)\sqrt{E^2-m^2\lambda(r)}}}.
\label{time}
\end{equation}
The minimum radius that can be reached by a test particle corresponds to
$r_{\rm min}>r_{\rm sing}$ for which $\lambda(r=r_{\rm min})=E^2/m^2$.
Note that $r_{\rm sing}$ corresponds to the space-time singularity,
{\it i.e.}, $\lambda(r=r_{\rm sing})=\infty$ and $R(r=r_{\rm sing})=0$.
On the other hand, classical massless particles with zero angular
momentum do not feel the repulsive gravitational potential due to
increasing $\lambda(r)$, and they reach the space-time singularity
in a finite time
\footnote{Another example of gravitational backgrounds, which repel
massive particles, are Type III supergravity walls \cite{CVETICGRIFF},
which are static, planar configurations interpolating between specific
$N=1$ supergravity vacua with negative cosmological constant (two
anti-deSitter vacua).  In this gravitational background, a massive
particle cannot reach the boundary of the anti-deSitter space-time on
one side of such a wall.  On the other hand, a classical massless test
particle with zero transverse momentum can reach this boundary.
This is the reason that on the boundary of the anti-deSitter vacuum
one has to specify the boundary conditions \cite{AIS}. Note, however,
that the boundary of the anti-deSitter space {\it is not singular}.}.

Purely electrically charged solutions have quantum numbers of elementary
string states,
thus indicating that they should be identified with elementary string
states \cite{Duff3}.
On the other hand, purely magnetically charged
solutions are not in the string spectrum, and should therefore be
viewed as solitonic configurations.
The nature of their charges indicates that they correspond to a hybrid
(``bound state'') of a Kaluza-Klein monopole and an H-monopole \cite{BANKS},
whose relative signs of the corresponding magnetic charges are opposite.
Note, that these configurations are different in nature from the BPS
$H$-monopoles\cite{HMON}.

Dyonic solutions cannot be identified with elementary string excitations
either, since there are no dyonic elementary string states.  We believe
that these configurations should be included in the spectrum of states
contributing to the  full string dynamics.  It would be important to find
out whether or not such configurations could arise from higher dimensional
soliton configurations as discussed extensively in Ref. \cite{HTII}
\footnote{Dyonic configurations may also have higher dimensional
embedding into generalised supersymmetric wave solutions.  Note that
a subset of regular dyonic solutions which correspond to
configurations with dyonic Kaluza-Klein charges ($P_1^{(1)}\ne 0$,
and $Q_2^{(1)}\ne 0$) \cite{CYKK} have an embedding into six-dimensional
supersymmetric Brinkman plane wave solutions \cite{GIBBP}.

For purely electrically charged solutions, such an embedding has been
given in Ref. \cite{BEHRNDT2}.  See also Ref. \cite{KALL2} and
references therein.}.

In the weak coupling limit, {\it i.e.}, $\phi_\infty\to -\infty$, and
at a generic point  of moduli space purely magnetically charged and
dyonic configurations decouple, while purely electrically charged ones
are light.  Note also that for purely magnetic [electric] singular
configurations the coupling becomes weak, {\it i.e.}, $\phi\to -\infty$,
[strong, {\it i.e.}, $\phi\to +\infty$,] near the naked singularity at
$r_{sing}$.  On the other hand, for dyonic singular configurations, with
relative signs of both of the corresponding charges opposite, the coupling
at the naked singularity is either weak  when ${\rm min}
[|{\bf P}^{(1)}_{1\,\infty}|,|{\bf P}^{(2)}_{1\,\infty}|]>{\rm min}
[|{\bf Q}^{(1)}_{2\,\infty}|,|{\bf Q}^{(2)}_{2\,\infty}|]$ or strong
when ${\rm min}[|{\bf P}^{(1)}_{1\,\infty}|, |{\bf P}^{(2)}_{1\,\infty}|]
< {\rm min}[|{\bf Q}^{(1)}_{2\,\infty}|, |{\bf Q}^{(2)}_{2\,\infty}|]$.

We would now like to turn to the role that such states may
play in enhancing the symmetries of vacua at special points of moduli
space.  For the example of the singular  solutions, presented above, the
allowed quantized magnetic and electric charge lattice vectors, as
determined \cite{SEN2} by the symmetries of the toroidally compactified
heterotic string \cite{NARAIN} and the DSZW quantization rule
\cite{DSZ,WITTENIII}, that could give rise to massless BH's correspond
to $\beta_1^{(1)}=-\beta_1^{(2)}=\pm 1$ and $\alpha_2^{(1)}=
-\alpha_2^{(2)}=\pm 1$, respectively.

The singular solutions which preserve $1\over 2$ of supersymmetries
correspond to purely, say, electrically charged solutions with
{\it two} choices for charge lattice vectors:
$\alpha^{(1)}_2= -\alpha^{(2)}_2=\pm 1$.  These two sets of solutions
become massless at the self-dual point of the ``one-torus'', {\it i.e.},
when $g_{22\,\infty}=1$.  Each of them belongs to a massless vector
super-multiplet, identified with the perturbative string states,
and form, together with the $U(1)^{(1)+(2)}$ gauge field
$(A^{(1)}_{\mu\,2}+A^{(2)}_{\mu\,2})/\sqrt 2$, the adjoint
representation of the non-Abelian $SU(2)^{(1)+(2)}_2$ gauge group,
thus enhancing the gauge symmetry from  $U(1)^{(1)}_2 \times U(1)^{(2)}_2$
to $U(1)^{(1)-(2)}_2 \times SU(2)^{(1)+(2)}_2$ at the self-dual
point of a one-torus.  In addition, there are purely magnetically
charged monopole configurations with $\beta_2^{(1)}=-\beta_2^{(2)}=\pm 1$,
discussed above, which along with a tower of the corresponding
$SL(2,Z)$ related dyonic states become massless at this point as
well, thus providing an explicit realization of the phenomenon,
studied by Hull and Townsend \cite{HTII}.

On the other hand, the singular solutions which preserve
$1\over 4$ of supersymmetries correspond to dyonic solutions with
electric and magnetic charges associated with two different one-tori
with four charge assignments $\beta_1^{(1)}=-\beta_1^{(2)}=\pm 1$ and
$\alpha^{(1)}_2=-\alpha^{(2)}_2=\pm 1$.  They become massless at the
self-dual point of the corresponding two-torus, {\it i.e.}, when
$g_{11\,\infty}=g_{22\,\infty}=1$.  Each of these configurations belongs
to a massless highest spin $3\over 2$ supermultiplet, which contains not
only massless vector fields (which combine with the $U(1)^{(1)+(2)}_1$
gauge field $(A^{(1)}_{\mu\,1} +A^{(2)}_{\mu\,1})/\sqrt 2$ and the
$U(1)^{(1)+(2)}_2$ gauge field $(A^{(1)}_{\mu\,2} +A^{(2)}_{\mu\,2})
/\sqrt 2$ to form the adjoint representation of the non-Abelian
$SU(2)^{(1)+(2)}_1 \times SU(2)^{(1)+(2)}_2$ gauge group), thus
contributing to the enhanced non-Abelian gauge symmetry from
$U(1)^{(1)}_1 \times U(1)^{(2)}_1\times U(1)^{(1)}_2 \times
U(1)^{(2)}_2$ to $U(1)^{(1)-(2)}_1\times U(1)^{(1)-(2)}_2 \times
SU(2)^{(1)+(2)}_1\times SU(2)^{(1)+(2)}_2$, but also four massless
spin $3\over 2$ states, which contribute to the enhancement of the
local supersymmetry.  Along with this set of dyonic states, there
is a tower of $SL(2,Z)$ related dyonic states, which also become
massless; at the self-dual point of the corresponding two-torus
one encounters a possibility of a new phase of the theory where,
along with a tower of massless vectors fields, a tower of massless
spin $3\over 2$  fields appears as well.

We would, however, like to caution that at the points of moduli space
with the enhanced symmetries the description of the effective field
theory breaks down due to the appearance of an infinite tower of new
massless fields, and thus the full string dynamics at such points is not
understood, yet. As for configurations belonging to the massless highest
spin $3\over 2$ supermultiplets, it may turn out that they correspond to
unstable states which decay into configurations belonging to the vector
supermultiplets.  All of these issues await further investigation.

\acknowledgments
The work is supported in part by U.S. Department of Energy Grant No.
DOE-EY-76-02-3071, the National Science Foundation Grant No. PHY94-07194,
the NATO collaborative research grant CGR 940870 and the National Science
Foundation Career Advancement Award  PHY95-12732. M.C. would like to thank
J. Polchinski, S.-J. Rey, A. Strominger and L. Thorlacius for useful
discussions, and C. Hull for helpful comments.

\end{document}